\RequirePackage{fix-cm}
\documentclass[smallextended]{svjour3}       
\smartqed  
\usepackage{graphicx}
 \usepackage{mathptmx}      
\usepackage{latexsym}

\usepackage{booktabs} 
\usepackage{array} 
\usepackage{paralist} 
\usepackage{verbatim} 
\usepackage{subfig} 
\usepackage{amsmath}                      	
\usepackage{amssymb}			
\usepackage{amsfonts}			

\begin{document}

\title{Bragg-Williams approximation for the dynamics of prey-predator biological associations
}


\author{E.M. De la Calleja$^{1}$, J.L.Carrillo$^{2}$, and I. Santamar\'{i}a-Holek$^{3}$}


\institute{$1$ Universidade Federal do Rio Grande do Sul, Caixa Postal 15051, 91501-970, Porto Alegre, RS, Brazil. \\
$2$ Instituto de F\'isica, Universidad Aut\'onoma de Puebla. Av. San Claudio y 18 sur, Edif. IF1, Ciudad Universitaria, C.P. 72570, Puebla, M\'exico. \\
$3$ Departamento de F\'isica, UMJ-Facultad de Ciencias, Universidad Nacional Aut\'onoma de M\'exico, Campus Juriquilla. Boulevard Juriquilla 3001, Quer\'etaro 76230, M\'exico.\\
              \email{corresponding author: elsama79@gmail.com} 
              }

\date{Received: date / Accepted: date}

\maketitle

\begin{abstract}
The dynamics of an association of interactive biological species is studied theoretically. We explore a mean field approximation to describe the temporal evolution of an ecological system with the basic prey-predator interspecies relation, as well as an approximation to introduce time correlations in the dynamics. We start by discussing the solution of the Volterra-Lotka model in a mean field approximation based in an analogy with the Weiss solution to the Ising model for ferromagnetic materials. In order to explore the effects of long-range time correlations, we describe the time evolution of the system within a kind of Bragg-Williams approximation. This approach allows us to evaluate a characteristic life-time of the ecosystem. This quantity could be very useful to discuss the time evolution of the system under a wide diversity of environmental conditions of the ecosystem which is not usually considered. We discuss the general trends of the temporal evolution of the association with some data from real ecosystems.

\keywords{Long-range time correlations \and prey-predator relationship \and mean field approximation \and Dynamics of biological associations}
\PACS{87.10.Hk\and 64.60.De \and 75.10-b \and 05.50.+q \and 05.65.+b}
\end{abstract}

\section{Introduction}\label{intro}
Population dynamics has been a along very long time a fundamental mathematical problem with strong implications in several fields of science and technology~\cite{Murray,Keshet,Abramson,Courchamp,Breno}.
Even presently, most of the fundamental aspects of the dynamics of systems of interacting species are still considered open problems.
Since the pioneering works of Lotka and Volterra~\cite{Lotka,Volterra} attempting a description based on a dynamical systems approach, there have been a large number of proposals trying to describe an ecosystem in terms of the internal and external variables of the biological associations, namely, the characteristics of the interactions among the species and the environmental conditions in which they are immersed.

There have been some other breakthrough proposals based on different approaches, for instance, statistical mechanics approaches~\cite{Kerner57,Kerner59}, stochastic and logistic approaches~\cite{Lande,Leslie}, and mean field approaches~\cite{Escudero,Adam,Bradshaw,Tibor,Tome,Nelson,Yoshida}.
Usually it is assumed that the biological associations of species are composed by a large number of populations of multiple different species, where the interactions among them depend on many biological and physical relations, as well as chemical reactions.

For most of these complex interactions no analytical expression is known.
In a real ecological system, as a result of these complex interactions the associations are able to organize hierarchically in many ways, but all of them are related to the primitive goals of survival and reproduction in a given environment.
In many cases the environment is the most relevant variable that determines the stability of the ecosystem.

It is usually assumed that the dynamics of the ecosystems could be described in an heuristic approximation taking into account its size, in terms of the number of individuals of the species, and the connection between them through complex interactions, usually parametrical introduced.
The time evolution of the system under these conditions generates some time patterns in the ecosystem.
One expects that the knowledge of correlations expressed in the hierarchical structures allows to describe the time evolution of the ecosystem in terms of the intrinsic relationships among the species, that express in some way their biological proximity.

In $1959$ E. Kerner {\em et al}~\cite{Kerner59} described in an statistical mechanics formalism, the dynamics of a biological association based on the Volterra-Lotka model. Authors discuss the equilibrium of the association considering the ensemble theory and determine the thermodynamic quantities as a function of the population.
They introduce the concept of eco-temperature which permit to calculate, free energy and heat capacity in terms of the population of the species.
In this way they discuss the characteristics of the steady states of the association.

In this report we focus our attention in an aspect of the dynamics of a biological association of species which usually is not addressed, this is the role of time correlations on the stability and dynamics of a close to equilibrium association of biological species.
By using the analogous number of coordination, we found the dynamical evolution of the ecological lattice as a function of the ecological coordination number and the number os species involved on the lattice.

To this end, we developed a mean field approximation based on a lattice model approach, and additionally propose and approximation to consider the effects of long range time correlations.
In the next section, we briefly review the essentials of the Ising lattice model, and introduce the lattice model to describe the biological association.
A mean field approximation based on an analogy with the Weiss approximation to solve the Ising Hamiltonian for ferromagnetic systems is discussed in section $3$.
To explore the effect of long range time correlations, in section $4$ an approximation based in an analogy with the Bragg-Williams is presented. Finally we make some remarks and comments.


\section{Ising model}\label{sec:2}

We present here a mean field model to describe the dynamics of a biological association with predator-prey interactions. We developed this lattice model taking advantage of the analogy between the mathematical expression of Issing Hamiltonian for ferromagnetic materials~\cite{Pathria}, with the rate equations which describe the time evolution of the populations of a biological association in the Volterra and Lotka model~\cite{Hernandez}.

Our lattice model allow us to describe a biological association focussing our attention in a particular pair of species immersed in an environment composed by the other species.
In other words, this approach allows us to collapse an association of biological species into a system formed by only three, where a pair of them are close biologically-wise, with a predator-prey interaction, and the third one varies slowly in time.

There exist a wide variety of systems whose dynamics can be addressed by using a lattice model approach.
Mean field approximation can be used to study the dynamics of a wide variety of interesting complex systems, such as: the behavior of social organization in a war~\cite{Naumis,Castello}, neural networks~\cite{Tracik}, and merging collective phenomena induced by local cellular automata~\cite{Grinstein}.
In the ecological case, the complex interaction of competition, cooperation or predation can be introduced by heuristic and phenomenological parameters that describe the rate of chance of the populations due to different interactions.

In the Ising model, named after Ernst Ising presented it in $1925$, first suggested by Wilhelm Lenz in $1920$~\cite{Brush}, each molecule in a lattice has a magnetic moment that can be oriented either up or down relative to the direction of an external applied field. Under these conditions the Ising hamiltonian include two terms, one of them considers the interaction of each magnetic moment with the external field, and the second includes the interaction at a given neighbor range.
In this way the Hamiltonian is given by

\begin{equation}
H_{I}=-\mu B\sum_{i}^{N}\sigma_{i} -J\sum_{i}^{N}\sum_{j}^{\gamma}\sigma_{j}\sigma_{i},
\label{eq1}
\end{equation}

where $\vec{B}$ is an external magnetic field, $\mu$ is the magnetic moment of the molecules, $N$ is the total number of molecules, $\sigma{i}$  (or $\sigma_{i}$)=+1 or an "up" spin and -1 for a "down" spin, and $\gamma$ represents the coordination number, i.e. the number nearest neighbors in a given site.
The geometrical aspects of the distribution of dipoles of the system are contained in this parameter.

Finally, \emph{J} stands for the exchange interaction~\cite{Pathria,Brush,Huang}.
Once introduced the Ising model, by taking into account the geometrical characteristics of a given lattice, the Hamiltonian can be cast into a form that incorporates these in the following form.

In the Ising model for a given configuration of the lattice, it is defined $N_{+}$ as the total number of sites of the lattice with spin-up magnetic moment, and $N_{-}$ as the total number of spin-down sites.
Considering only the nearest neighbors, the couples that can be formed from the spins can be only of three types (+,+), (-,-) and (+,-).
The relations that describe the exchange between the sites (+) and (-) are

\begin{equation}
\begin{array}{c}
\gamma N_{+} = 2N_{++} + N_{+-},\\
\gamma N_{-} = 2N_{--} + N_{+-},\\
N= N_{+} + N_{-}.
\end{array}
\label{eq2}
\end{equation}

By using these relations, the Ising hamiltonian can be cast into

\begin{equation}
\begin{array}{c}
H_{I} = -J(N_{++}+N_{--}-N_{+-})-\mu B(N_{+}-N_{-},\\
H_{I}=-4 J N_{++} + 2(\gamma J-\mu B) N_{+} + (\mu B-\gamma J/2)N,
\end{array}
\label{eq3}
\end{equation}

whose solution in terms of the number of pairs (+,+), (+,-) and (-,-) has been extensively discussed~\cite{Pathria,Huang}.
In $1907$ Pierre Weiss proposed an approximation to solve the Ising model.
By considering that the energy of an individual atom in a given configuration of the system, is determined by the average energy rather than by the fluctuating value of the neighboring atoms, he proposed to replace the sum over the nearest neighbors by its mean value.
It means that the Hamiltonian given by the equation (\ref{eq1}) can be written as

\begin{equation}
H=-\sum_{i}^{N}\mu_{iz}(B_{z} + B_{mol}),
\label{eq4}
\end{equation}

where $B_{mol}=J\sum_{i}^{\gamma}\mu_{jz}$. Thus, by means of the Weiss approximation, the many magnetic moment system has been cast into a system formed by independent magnetic moment under the influence of the total field,

\begin{equation}
H=-\sum_{i}^{N}(B_{tot})\mu_{iz}.
\label{eq5}
\end{equation}

$B_{tot}=B_z + B_{mol}$ qualitatively describes well the ferromagnetic-paramagnetic transition, however, it fails in predicting the scaling behavior of the physical quantities in the critical region. The physical reason of this fail is that the mean field approximation does disregard fluctuations in the system.

\subsection{The ecosystem lattice model}\label{sec:1}
We construct our lattice model for the biological association by developing expressions analogous to relations (\ref{eq2}) for a prey-predator system.
It is convenient to remark here the analogy between the mathematical structure of the Ising model Hamiltonian and the equations of the Lotka-Volterra model.
We recall that the description of the time evolution of the populations in an association of biological species in a traditional rate equations framework is based on expressions with the form \cite{May},

\begin{equation}
\frac{dP_{i}(t)}{dt}=F_{i}(p_{1}(t),p_{2}(t),...,p_{i}(t)),
\label{eq6}
\end{equation}

where $F_{i}$ is in general, some nonlinear function of the populations $p_{i}$, of the \emph{i-th} species. In this framework, one of the first models to analyze the dynamics of interactive biological species by joining the scheme developed by Vito Volterra~\cite{Vol28,Vol31,Vol37} and that proposed by Alfred J. Lotka~\cite{Lot20,Lot10,Lot56}, in order to describe autocatalytic reactions.

The Lotka-Volterra model \emph{(LVM)} describes time evolution of the biological association based on a set of non linear rate equations. This model predicts a cyclic behavior of the populations of a given number of species. It has been considered that this is a kind of primitive model of a biological association.

Depending on the initial conditions, these equations describe a whole family of curves in the phase space.
The \emph{LVM} initially proposed to describe a couple of species, has been generalized to describe $\gamma$-species, becoming the corresponding equations in the form

\begin{equation}
\frac{dN_{i}}{dt}= \pm\sum_{i}^{\gamma}\epsilon_{i}N_{i} + \beta_{i}^{-_{1}}\sum_{j}^{\gamma}\alpha_{ji}N_{j}N_{i}.
\label{eq7}
\end{equation}

Where the parameters $\epsilon_{i}$ are the rates of growth of the \emph{i-th} specie, which can be can be positive or negative depending of the role of that specie.
If $\epsilon_{i}$$>$0 the $\emph{i-th}$ population grows following the Malthus law~\cite{Malthus} and if $\epsilon_{i}$$<$0, and $N_{j}$=0 the population of the $\emph{i-th}$ specie decays exponentially.

The binary encounters between individuals of the populations $\emph{i-th}$ and $\emph{j-th}$ is the quadratic term given by $N_{i}N_{j}$ and its effect is governed by the coefficients $\alpha_{ji}$ which define the rate at which $N_{i}$ increases or decreases due to the encounter with the other specie.
When the encounters between the members of the species \emph{i} and \emph{j} occur, they lead to the situation where one of the species wins and the other looses.
It means that $\alpha_{ij}$=$-\alpha_{ji}$.
Coefficient $\beta_{i}$ is the so called Volterra's number or equivalence of the ratio of \emph{i}'s losers (or gainers) per unit time to \emph{j}'s gainers (or losers) expressed as $\beta_{i}^{-1}/\beta_{j}^{-1}$~\cite{Kerner57}.

\subsection{Lattice model for a prey-predator system}
Our lattice model for a biological association is constructed taking advantage of the mathematical structural similarity between the rate equations of the \emph{LVM} with the Ising Hamiltonian.

We assume that the dynamics of the total population is governed by the interactions preys-predators populations, where the prey and predator population is labeled by $N_{p}$ and $N_{d}$ respectively.
Of course in this simplified approximation, aspects like sex and age are disregarded to construct the basic equations of the model.
At a given time the expressions analogous to equations (\ref{eq2}) for biological association of species of a given role, i.e. prey or predator, can be expressed for prey role,

\begin{equation}
\begin{array}{c}
\gamma N_{p}=\xi_{p}N_{p}N_{p}-(1-\phi)N_{p}N_{d}=\gamma N_{p}=\xi_{p}N_{p}N_{p}-\lambda_{1} N_{p}N_{d}=\\
\gamma N_{p}=\xi_{p} N_{pp}-\lambda_{1} N_{pd},  \\
\end{array}
\label{eq8}
\end{equation}

and for the predator case,

\begin{equation}
\begin{array}{c}
\gamma N_{d}=\xi_{d}N_{d}N_{d}+(1-\phi)\nu N_{d}N_{p}=\gamma N_{d}=\xi_{d} N_{d}N_{d}+\lambda_{2} \nu N_{d}N_{p}= \\
\gamma N_{d}=\xi_{d} N_{dd}+\lambda_{2} \nu N_{dp},
 \end{array}
 \label{eq9}
\end{equation}

taking into account that the total population is the sum of preys and predators populations as,

\begin{equation}
N_{T}=N_{d}+N_{p}.
\label{eq10}
\end{equation}

The $\gamma$ parameter is in the ecological association the analogous to the coordination number, and is a measure of the biological neighborhood. $\xi_{i}$ is the parameter that describes the encounters by pairs or groups at a given role of species, i.e. prey \emph{(p}) or predator \emph{(d)}.

$\lambda_{i}$ with $i=1,2$ is related to the fraction of prey caught per predator per unit time, it means, the hunting efficiency of the predator and the ability of the prey to scape.
If $\phi=1$ in the interval [0,1] it implies that there are not ideal encounters between a prey and a predator that favoring the predators.
If $\phi$=0 the encounters of species is ideal, and decrease the number of preys, because it indicates that the prey was successfully hunted by the predator.
In some way, the preys eaten are converted in newborn predators, $\nu$ represents the conversion factor.

The temporal evolution of the biological association for $\gamma$ species can be expressed in a form analogous to equation (\ref{eq7}) in an expanded form as,

\begin{equation}
\frac{dN_{T}}{dt}=\sum_{i}^{\gamma}([{\epsilon_{p}N_{p}+\epsilon_{d}N_{d}]+\frac{1}{\beta_{p}}[\alpha_{pp}N_{p}N_{p}+\alpha_{dd}N_{d}N_{d}-2\alpha_{pd}N_{p}N_{d}}]).
 \label{eq11}
\end{equation}

According to relations (\ref{eq8}),(\ref{eq9}) and (\ref{eq10}) we can obtain each term of the equation (\ref{eq11}) in terms of the number of preys using the next relations,

\begin{equation}
\begin{array}{c}
N_{dp}=\frac{\xi_{p}}{\lambda_{1}}N_{pp}-\frac{\gamma}{\lambda_{1}}N_{p}, \\
N_{dd}=\frac{\gamma}{\xi_{d}}N_{T}-\frac{\gamma}{\xi_{d}}(1+\frac{\lambda_{2}\nu}{\lambda_{1}})N_{p}-\frac{\lambda_{2}\nu\xi_{p}}{\lambda_{1}\xi_{d}}N_{pp},
\end{array}
\label{eq12}
\end{equation}

The rate equation for the total population as (\ref{eq11}) can be written explicitly in terms of preys in the form,

\begin{multline}
\frac{dN_{T}}{dt}=\Psi =(\epsilon_{d} + \frac{\gamma\alpha_{dd}}{\beta_{p}\xi_{d}})N_{T}+ (\epsilon_{p}-\epsilon_{d})N_{p} +\frac{\gamma}{\beta_{p}} \Bigg[ \frac{2\alpha_{pd}}{\lambda_{1}} - \frac{\alpha_{dd}}{\xi_{d}}(1-\frac{\lambda_{2}\nu}{\lambda_{1}})\Bigg] N_{p}\\
- \Bigg[ \frac{\xi_{p}}{\beta_{p}\lambda_{1}} (2\alpha_{pd} + \frac{\alpha_{dd}\lambda_{2}\nu}{\xi_{d}})-\frac{\alpha_{pp}}{\beta_{p}}\Bigg] N_{pp}.
 \label{eq13}
\end{multline}

Also we can express the rate equation (\ref{eq11}) in terms of predator population, leading to,

\begin{multline}
\frac{dN_{T}}{dt}=\Psi=(\epsilon_{p}+\frac{\gamma\alpha_{pp}}{\beta_{d}\xi_{p}})N_{T}+ (\epsilon_{d}-\epsilon_{p})N_{d} + \frac{\gamma}{\beta_{d}} \Bigg[ \frac{\alpha_{pp}}{\xi_{p}} (\frac{\lambda_{1}}{\lambda_{2}\nu}-1) - \frac{2\alpha_{pd}}{\lambda_{2}\nu}\Bigg] N_{d}\\
+ \frac{1}{\beta_{d}} \Bigg[ {\alpha_{dd}}+\frac{\xi_{d}}{\lambda_{2}\nu}(2\alpha_{pd}-\frac{\alpha_{pp}\lambda_{1}}{\xi_{p}})\bigg] N_{dd}.
 \label{eq14}
\end{multline}

Equations (\ref{eq13}) and (\ref{eq14}) represents the temporal evolution of the total population taking into account $\gamma$ species, and its mathematical form is similar to equation (\ref{eq1}) for the ferromagnetic system.

\section{Mean field approximation}
The similarity between the mathematical structure of the Lotka-Volterra equation for total population of an association of biological species (equation \ref{eq7}), with the Ising model established by equation (\ref{eq4}), suggests the possibility of constructing a mean field approximation, analogous to equation (\ref{eq5}), to describe the evolution of the total population of the biological association by mean field theory.
The rate equation for the total population may be written in the form,

\begin{equation}
\frac{dN_{T}}{dt}=\sum_{i}^{\gamma}\epsilon_{i}N_{i}+\frac{1}{\beta_{i}}\sum_{ij}^{\gamma}\alpha_{ji}N_{j}N_{i},
\label{eq15}
\end{equation}

the sum is in the subindex $<ij>$ is over the nearest-neighbors. One may define a mean field parameter for the \emph{i-th}-specie in the form

\begin{equation}
T_{eff_{i}}=\sum_{j}^{\gamma}\alpha_{ji}N_{j}.
\label{eq16}
\end{equation}

Considering only the prey-predator relationship, it is possible to define a mean field parameter for the total number of species.
This parameter would include the interaction between nearest-neighbors for the \emph{i-th} specie.
The temporal evolution of the the total population in terms of the effective parameter would be given by,

\begin{equation}
\frac{dN_{i}}{dt}=\sum_{i}^{N}[\epsilon_{i}+\beta_{i}^{-1}T_{eff_{i}}]N_{i}.
\label{eq17}
\end{equation}

By considering the definition of the mean field term, for the \emph{i-th} specie we have,
\begin{equation}
T_{cm_{i}}=\epsilon_{i}+\beta_{i}^{-1}T_{eff_{i}}.
\label{eq18}
\end{equation}

In terms of the above definition, the rate equation for the total population in the biological association adopts the form,

\begin{equation}
\frac{dN_{T}}{dt}=\sum_{i}T_{cm_{i}}N_{i}.
\label{eq19}
\end{equation}

The above procedure, allow us to express equation (\ref{eq16}) for the two specific roles, it means in terms of $N_{d}$ and $N_{p}$ parameters, therefore can it be written as,

\begin{equation}
\begin{array}{c}
T_{eff_{p}}=\alpha_{dp}N_{d}+\alpha_{pp}N_{p},\\
T_{eff_{d}}=\alpha_{pd}N_{p}+\alpha_{dd}N_{d}.
\end{array}
\label{eq20}
\end{equation}

By using the relations (\ref{eq8}), (\ref{eq9}) and (\ref{eq10}) we obtain just equation (\ref{eq13}) or (\ref{eq14}).

\subsection{Weiss approximation}
As far as now we have defined a lattice model for the biological association, the parameter $\gamma$ represents the biological proximity, in analogy with the coordination number in the Ising model.
In a real ecosystem not all the species interact among them simultaneously, thus, one might think that even though they share a common territory, some species are not noticed by others.
Taking \emph{$\kappa$} species that interact each other by the prey-predator relation, the biological proximity represented by the $\gamma$ parameter allows to assume that

\begin{equation}
\sum_{j}^{N}\frac{\alpha_{ji}N_{j}}{\beta_{i}}\sim\sum_{j}^{\kappa}\frac{\alpha_{ji}N_{j}}{\beta_{i}},
\label{eq21}
\end{equation}

with $\kappa<N$. By using the definition of the parameters $\epsilon_{eff_{i}}$ and $\epsilon_{cm_{i}}$, the rate equations which describes the temporal evolution of the \emph{i-th} specie and the temporal evolution of the total population of the association can be cast into

\begin{equation}
\frac{dN_{i}}{dt}=[\epsilon_{i}+\sum_{j}^{\kappa}\alpha_{ji}N_{j}]N_{i}(t),
\label{eq22}
\end{equation}

\begin{equation}
\frac{dN_{T}}{dt}=[\epsilon+\sum_{i}^{N}{\sum_{j}^{\kappa}\alpha_{ji}N_{j}}]N_{T}(t).
\label{eq23}
\end{equation}

By introducing a mean field parameter for the \emph{i} species, and the mean field for the total population one obtains

\begin{equation}
\Omega_{T}(k)=\sum_{i}^{N}(\sum_{j}^{\kappa}\alpha_{ji}N_{j}),
\label{eq24}
\end{equation}

\begin{equation}
\Omega_{i}(k)=\sum_{j}^{\kappa}\alpha_{ji}N_{j}.
\label{eq25}
\end{equation}

The temporal evolution of the total population and for the \emph{i-th} population are then given by

\begin{equation}
\begin{array}{c}
dN_T/dt = \Bigg[ \epsilon+\Omega_{T}(k)\Bigg] N_{T}(t),\\
dN_i/dt = \Bigg[ \epsilon_{i}+\Omega_{i}(t)\Bigg] N_{i}(t).
\end{array}
\label{eq26}
\end{equation}

By integrating (\ref{eq26}) one obtains the expression or $N_{T}$ as,

\begin{equation}
\begin{array}{c}
N_{T}(t)=N_{T}(t_{0})\Bigg[ \exp[\epsilon+\Omega_{T}(k)] \Bigg] (t-t_{0}),\\
N_{i}(t)=N_{i}(t_{0})\Bigg[ \exp[\epsilon_{i}+\Omega_{i}(k)] \Bigg] (t-t_{0}).
\end{array}
\label{eq27}
\end{equation}

In this way, with $(t-t_{0})$

\begin{equation}
\begin{array}{c}
lnN_{T}(t)=lnN_{T}(t_{0})\Bigg[ (\epsilon+\Omega_{T}(k))\Delta t\Bigg], \\
lnN_{i}(t)=lnN_{i}(t_{0})\Bigg[ (\epsilon_{i}+\Omega_{i}(k))\Delta t\Bigg].
\end{array}
\label{eq28}
\end{equation}

Assuming that $\Delta t$ is a small quantity compared with the total time of observation, we obtain the following expressions for the total and \emph{$i-th$}specie mean fields.

\begin{equation}
\widetilde{\Omega_{T}(k)}=\frac{lnN_{T}(t)-lnN_{T}(t_{0})}{\Delta t} - \varepsilon_{T} = \frac{dlnN_{T}(t)}{dt} - \varepsilon_{T},
\label{eq29}
\end{equation}

\begin{equation}
\widetilde{\Omega_{i}(k)}=\frac{lnN_{i}(t)-lnN_{i}(t_{0})}{\Delta t} - \varepsilon_{i} = \frac{dlnN_{i}(t)}{dt} - \varepsilon_{i}.
\label{eq30}
\end{equation}

\subsection{Long range correlations}
Following the pioneering work of Gorsky~\cite{Gorsky} about the order-disorder transition in binary alloys, Bragg-Williams ($1934$-$1935$)~\cite{Bragg34,Bragg35,Williams} introduced the concept of long-range correlation. The idea of the Bragg-Williams approximation is that the energy of a single atom in a given system is rather determined by the average order degree prevalent for the total system than by the fluctuation in the local configuration of the atoms.

It has been discussed that the approximation formulated by Gorsky and the formulated by Bragg and Williams are physically equivalent~\cite{Fowler}. Behind this approximation also lays the origin of the mean field theory approach developed by Weiss~\cite{Pathria}.

The Bragg-Williams approximation for the ferromagnetic systems expresses the Hamiltonian of system in terms of the long-range \emph{(L)} and the short-range ($\sigma$) correlations.
In the Ising model the energy of the system depends on the number of sites with spin up ($N_{+}$) and pair of nearest-neighbor ($N_{++}$) (see equation (\ref{eq2})).

The relation $\frac{N_{++}}{\gamma N/2}$ is a measure of the short-range order and gives us the fraction of its nearest-neighbors with spin up. $N_{+}/N$ is a measure of long-range order that considered the entire lattice of all spins must be up. We follow the same considerations proposed in the Ising model~\cite{Huang}, for that we define the long-range order in terms of \emph{(L)} and the short-range order in terms of $\sigma$ as $N_{+}/N$=$(L+1)/2$ and $\frac{N_{++}}{\gamma N/2}=(\sigma+1)/2$. Of course the numerical value of these correlations always, are in the range $(-1\leq\emph{L}\leq1)$ and $(-1\leq\sigma\leq1)$.

The Bragg Williams approximation take into account that $\frac{N_{+}}{N}\approx \frac{N_{++}}{\gamma N/2}$, this based in the assumption that "there is no short-range order apart from that which follows from long-range order"~\cite{Bragg35}. The Ising hamiltonian in terms of Bragg-Williams approximation in terms of \emph{(L)} can be expressed in the form

\begin{equation}
H_{I}=-\frac{1}{2}JN\gamma\bar{L}^{2} - \mu BN\bar{L}.
\label{eq31}
\end{equation}

\subsection{Effects of long time correlations}

In order to express the corresponding rate equations for the total population of the biological association in the Bragg-Williams approximation, it is necessary to introduce the short and long-range correlations into the expressions analogous to the Ising Hamiltonian, it is done by the following procedure.

\begin{equation}
N_{p}=N_{T}\frac{(L+1)}{2},
\label{eq32}
\end{equation}
and
\begin{equation}
N_{pp}=N_{T}\frac{\gamma}{4}(\sigma+1)=N_{T}\Bigg[\frac{\gamma L^{2}}{8}+\frac{\gamma}{4}(L+1)-\frac{\gamma}{8}\Bigg],
\label{eq33}
\end{equation}

with the obvious meaning of $N_{p}$, $N_{pp}$ and $\gamma$. By the substitution of these expressions in the rate equation for the total population of the ecosystem. In a ferromagnetic material the Bragg-Williams Hamiltonian describes the energetic state of the system, following with our analogy in the case of an biological association the corresponding Hamiltonian is the rate equation for the total population, in this case larger changes in $N_{T}$ should be interpreted as stages of high energy ($\Psi$) in the system. Expressing our lattice model in terms of long-range correlations, we have

\begin{multline}
\frac{dN_{T}}{dt}= \Psi = N_T L^2 \frac{\gamma}{\beta_p} \Bigg[ \frac{\alpha_{pp}}{8} - \frac{\xi_p}{\lambda_1} \Bigg( \frac{\alpha_{pd}}{4}-\frac{\alpha_{dd}\lambda_2 \nu}{8\xi_{d}}\Bigg) \Bigg]+ N_{T}L(\frac{\epsilon_p - \epsilon_d}{2})+
\\
\qquad \qquad N_T L \frac{\gamma}{\beta_p\lambda_{1}} \Bigg\{ \Bigg[ \alpha_{pd} (1- \frac{\xi_p}{2})\Bigg] + \frac{1}{\xi_d} \Bigg[ \frac{\alpha_{dd}\lambda_{2}\nu}{2}(1-\frac{\lambda_1}{\lambda_2\nu}-\frac{\xi_{p}}{2}) + \frac{\alpha_{pp}\lambda_{1}\xi_{d}}{4}\Bigg]\Bigg\} +
\\
N_{T}\Bigg[ (\frac{\epsilon_p - \epsilon_d}{2})+\epsilon_{d}\Bigg] + N_{T}\frac{\gamma}{\beta_p} \Bigg[ \frac{3\alpha_{pp}}{8} +
\frac{\alpha_{dd}}{\xi_d} \Bigg( \frac{1}{2} + \frac{\lambda_{2}\nu}{2\lambda_{1}}-\frac{\xi_{p}\lambda_{2}\nu}{\lambda_{1}}\Bigg) + \frac{\alpha_{pd}}{\lambda_{1}}(1-\frac{\xi_{p}}{4}) \Bigg].
\label{eq34}
\end{multline}

\section{Results and conclusions}
Now we explore on the basis of these equations, the influence of the long-range time correlations on the time evolution of total population of the association. In this procedure we take the coordination fixed values of $\beta_{p}=0.006456$~\cite{Hernandez}, $\alpha_{pd}=0.05$ and $\nu=0.05$. The other values were determined using relations (\ref{eq8}), (\ref{eq9}) and (\ref{eq10}). For example if the interaction between preys does not exit ($\alpha_{pp}=0$), it determines that the encounters, between them also does not exists for that $\xi_{p}=0$. It affect the effective hunting or ability to scape represented by $\lambda_{i}$. In the next table there appears a list of values of these parameters we used in our calculation.

\begin{table}[h!]
\centering
\begin{tabular}{|l|l|l|l|l|l|}
  \hline
  $\alpha_{pp}$ & $\alpha_{dd}$ & $\lambda_{1}$ & $\lambda_{2}$ & $\xi_{p}$ & $\xi_{d}$ \\ \hline
  0 & 1 & 1 & 0 & 0 & 1 \\ \hline
  0.2 & 0.2 & 0.8 & 0.8 & 0.2 & 0.8 \\ \hline
  0.6 & 0.4 & 0.6 & 0.4 & 0.6 & 0.4 \\ \hline
  0.8 & 0.8 & 0.2 & 0.2 & 0.8 & 0.2 \\ \hline
  1 & 0 & 0 & 1 & 1 & 0 \\ \hline
  \end{tabular}
  \caption{Values of $\alpha_{pp}$, $\alpha_{dd}$, $\lambda_{1}$, $\lambda_{2}$, $\xi_{p}$ and $\xi_{d}$ used to obtain numerical results of equation ~(\ref{eq34}).}
\end{table}\label{Tab-1}

We have considered the following experimental data from the Serengeti ecosystem in Tanzania and Kenya, East Africa. We consider $10$ species of large carnivores, and $28$ different species of preys~\cite{Preda-2}. The carnivore animals in consideration are: lion, hyena, leopard, cheetah, wild dog, silver-backed jackal, serval, caracal, wild cat and golden jackal~\cite{Suplementary}. Some of the preys in the ecosystem are: wildebeest, zebra, gazelle, oribi, buffalo or giraffe~\cite{Preda-2}.

We start our analysis calculating the change in time of the energy $\Psi$ of our ecological lattice, for several values of $\gamma$, which represents the biological neighborhood. $\gamma$ can be adopt values of $1-5$ taking into account the kilograms that those predators can eat from different preys and the average number of predators that can be hunt or eat the same prey. By considering the average days for gestational of five predators (lion, hyena, wild dog, leopard and cheetah) and some preys (wildebeest, zebra, gazelle, oribi and buffalo), we use the next relation to determine the growth rate of preys and predators

\begin{equation}
\varepsilon_{i}=\frac{1}{\tau_{i}}=\sum_{i=1}^{\gamma}\frac{1}{\tau_i},
\label{eqt}
\end{equation}

where the sum is over all the neighborhood species for preys and predators respectively. By equation (\ref{eqt}) and consider a total number of $1.1$ predators per $km^{2}$, and a density of $3$ preys per $km^{2}$~\cite{Predators}, $\epsilon_{p}=0.34$ preys per month and $\epsilon_{d}=0.294$ predators per month taking into account and average gestational time.

\begin{figure}[!h]
\begin{center}
 \includegraphics[width=0.7\textwidth]{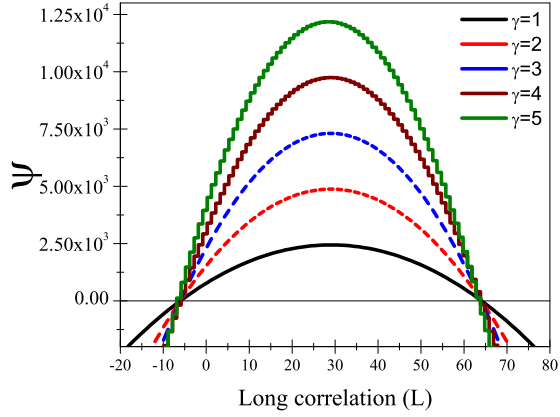}
 \caption{The change in time of the total population as a function of the long range time correlation \emph{L} in units per month, for various values of the biological association $\gamma$ with initial value of $N_{T}=4.1$ per $km^{2}$.}
 \label{Fig-1}
\end{center}
\end{figure}

\begin{figure}[!h]
\begin{center}
 \includegraphics[width=0.7\textwidth]{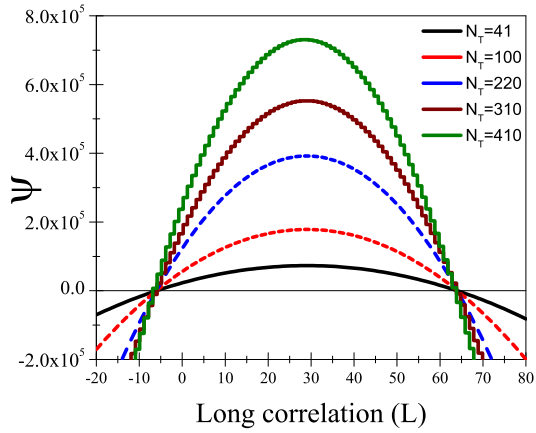}
\caption{The change in time of $\Psi$ as a function of of \emph{L} in units per month with $\gamma$=3 for different values of $N_{T}$ per $km^{2}$}
\label{Fig-2}
\end{center}
\end{figure}

By solving numerically the rate equations and taking a variational space from $1km^{2}$ to $100km^{2}$, we obtain the time evolution of the system. In Fig.(\ref{Fig-1}) is plotted the "energy" $\Psi$, as a function of correlations \emph{(L)} in units per month, for values of $0.2<\alpha_{pp}<0.8$. When $\alpha_{ii}$ takes extreme values, the roots are complex. The "energy" $(\Psi)$ of the system increases as a function of $\gamma$, which represents the ecological association of species similar to the coordinate number on the Ising model.
On the other hand, if one keep constant $\gamma=3$ and changes $N_{T}$ in the interval (41-410) $km^{2}$, the behavior of the temporal rate of change of the total population is shown in the figure (\ref{Fig-2}).

Notice that for all the the $\gamma$ values, the largest changes in the total population of the association occur when the long time correlation attains a value around $30$. Obviously, the system is more stable in its total population when the biological proximity has its lowest value. In Fig.(\ref{Fig-2}) we obtain that the "energy" $(\Psi)$ of the system decrease as a function of the number of populations that interact in the ecological system.

\begin{figure}[!h]
\begin{center}
\includegraphics[width=8cm]{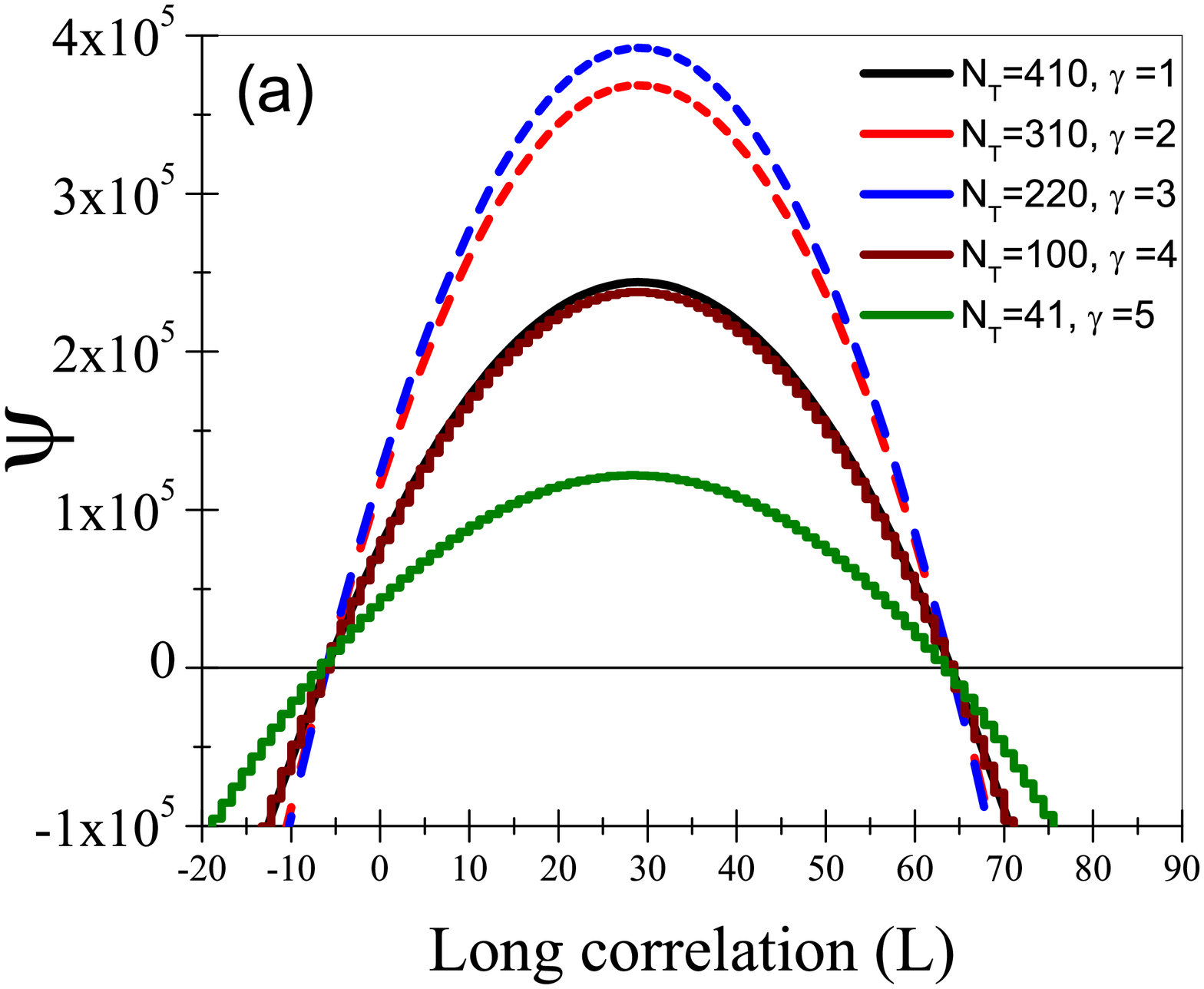}
\includegraphics[width=8cm]{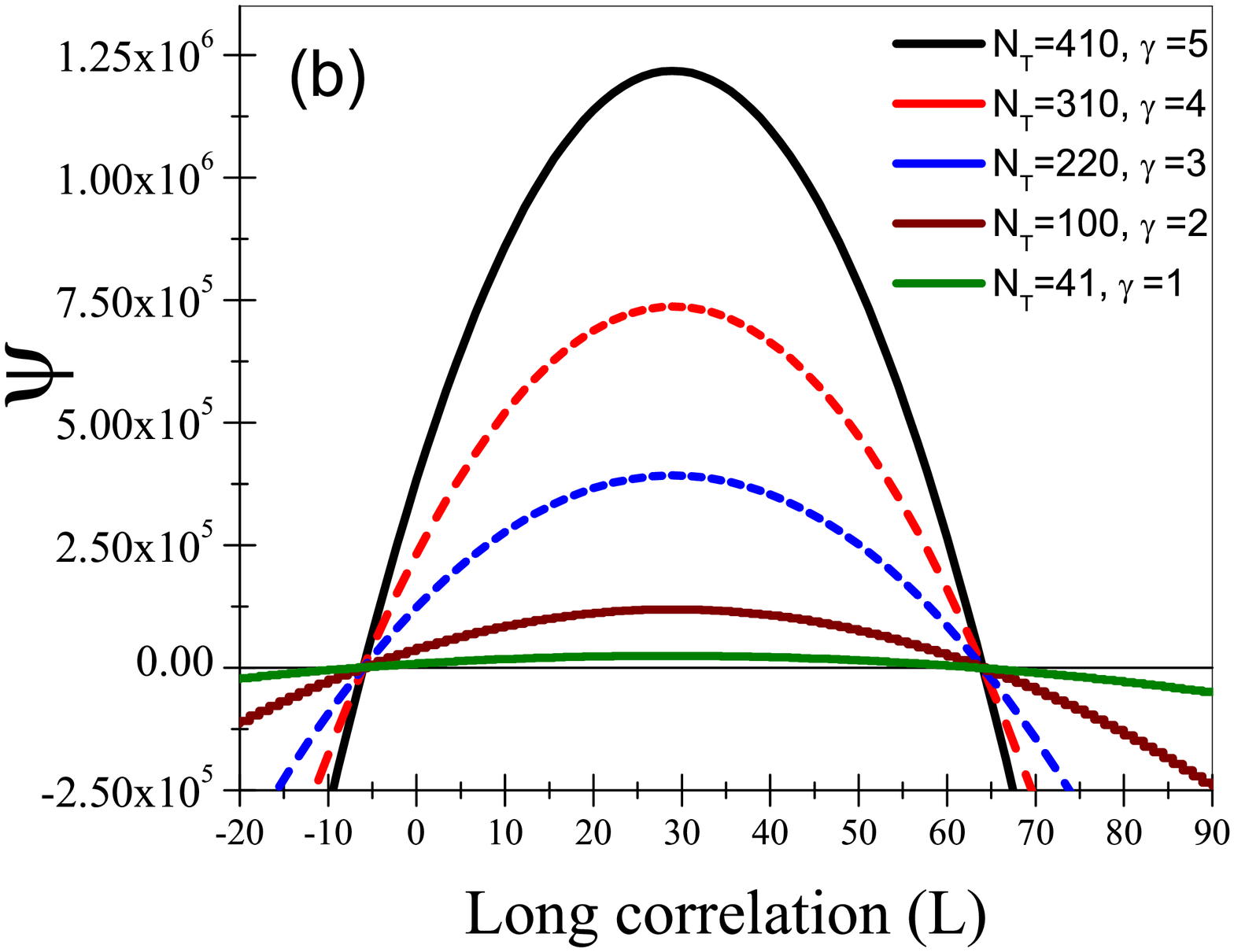}
\end{center}
\caption{\emph{(a)} show the simultaneous change of $\Psi$ as a function of \emph{L} in units per month from $N_{T}$ from (410-41) at the same time with change of $\gamma$ from 1 to 5; \emph{(b)} show the behavior of $\Psi$ as a function of \emph{L} in units per month, for the same variation of $N_{T}$ and for $\gamma$ changing from 5 to 1.}
\label{Fig-3}
\end{figure}

If simultaneously $\gamma$ and $N_{T}$ are changed, one obtains the behavior shown in Fig.(\ref{Fig-3}).
The system is very sensitive to changes in $\gamma$ and $N_{T}$.
By decreasing $N_{T}$ and increasing $\gamma$, the maximum of the changes in the total population is obtained as a function of long correlations (L) is shown in Fig.~\ref{Fig-3}\emph{(a)}.
The intermediate values of $N_{T}$ and $\gamma$ obtain larger energy values.
For example for $N_{T}=220$ and $\gamma=3$ the energy of the system attains its maximum value.
And for $\gamma=5$ and $N_{T}=41$ the system energy reaches its minimum value. Fig.~\ref{Fig-3}\emph{(b)} shows how decreasing values of $N_{T}$ and decreasing values of $\gamma$ affects the time evolution of the association.
In Fig.~\ref{Fig-3}\emph{(b)} can be corroborated the heavy influence of $\gamma$ on the dynamics in the systems

\begin{figure}[!h]
\begin{center}
 \includegraphics[width=0.7\textwidth]{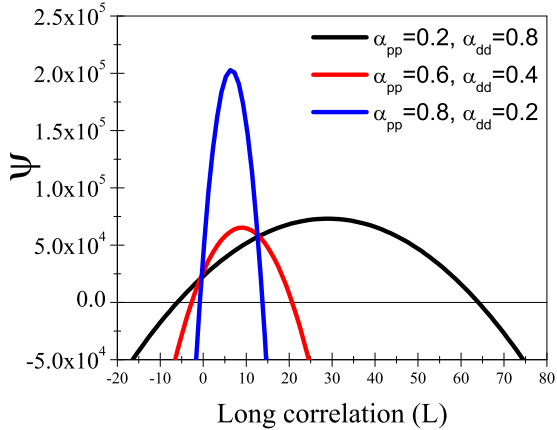}
 \caption{The change in time of the total population as a function of the long range time correlation $L$ in units per month for various values of the $\alpha$ for $N_{T}=41$ and $\gamma=3$.}
 \label{Fig5}
\end{center}
\end{figure}

In Fig.(\ref{Fig5}) it is shown the behavior of $\Psi$ as a function of long correlation for different values of $\alpha$ according to the table (1). The effect of the long time correlation on the time evolution of the system is stronger, and the system is more sensitive in a more localized interval of values of \emph{L} in units per month. This is more noticeable for increasing values of $\alpha_{pp}$.

\section{Comments and remarks}\label{sec:5}
It is well known that the main limitation of the mean field approaches is that the effect of the fluctuation is disregarded.
Of course this is a serious obstacle to describe the system near critical conditions.
However, even far from the critical conditions, the fluctuations can play an important role in the dynamics of the biological association.
Here we have shown that by means a mean field approximation it is possible to analyze the dynamics of a predator-prey association, and was possible to include the interaction between nearest-neighbors for the \emph{i-th} specie. We found the analytical expression for the temporal evolution of the the total population in terms of the effective parameters which were correlated with some experimental data of 10 different species of predators and 28 preys interacting with those population.

In these conditions the mean field parameters con be time dependent quantities.
Under the proper conditions, this lattice model solved through a mean field approach constructed as an analogy with the Weiss approximation, allows us to describe the total and \emph{$i-th$}specie mean fields, to develop an analytical expression for the "energy" of the time evolution of an ideal predator-prey association.

We have constructed a novel approach, the Bragg-Williams approximation for the lattice model corresponding to the Ising model of a biological association, moreover, we have shown that it is possible to mount the qualitative description of the fluctuations given by the Bragg-Williams approximation on the mean field Weiss model constructed for a biological association. The procedure is internally consistent and qualitatively predicts the trends that one should expect about the role of the correlations.

The general trends shown in figures is that when the system contains short time correlations, the rate of change of  the total population is small as well. This is due to the influence of the correlation is very local in time. On the other extreme, when the correlation time is large compared to the characteristic time of variation of the populations, the system reaches the maximum value of "energy" related to the rate of change of the total population and the coordination number $\gamma$.


%
%

\begin{acknowledgements}
E.M.C.M. Bolsista CNPq. Partial financial support by CONACyT and VIEP BUAP, M\'exico. Partial support by Red Tem\'atica de Materia Condensada Blanda, CONACyT M\'exico.
\end{acknowledgements}



\end{document}